\documentclass[aps,pre,twocolumn,showpacs]{revtex4-1}
\usepackage{epsfig,graphicx}
\usepackage{natbib}
\usepackage[nooneline]{subfigure}

\begin{document}


\title{Occupation times on a comb with ramified teeth}


\author{A. Rebenshtok, E. Barkai}
\affiliation{Department of Physics, Bar Ilan University, Ramat-Gan 52900 Israel}

\email{barkaie@biu.ac.il}


\date{\today}

\begin{abstract}
We investigate occupation time statistics for random walks on a comb with ramified teeth. This is achieved through the relation between the occupation time and the first passage times. Statistics of occupation times in half space follows Lamperti's distribution, i.e. the generalized arcsine law holds. Transitions between different behaviors are observed, which are controlled by the size of the backbone and teeth of the comb, as well as bias. Occupation time on a non-simply connected domain is analyzed with a mean-field theory and numerical simulations. In that case, the generalized arcsine law isn't valid. 
\end{abstract}


\maketitle

\section{Introduction}

The occupation time distribution in half space for Brownian motion on a one dimensional infinite line is well known. The distribution converges to the arcsine law found by P. L\'evy \cite{r5}. A general goal of this article is to find the deviation from the arcsine law for a comb system. 

The comb is a simplified model for various types of natural phenomena which belong to the loopless graphs category. Two examples are spiny dendrites of neuron cells \cite{r1212} and dendronized polymers \cite{r1213}. The comb model consists of a backbone and teeth (see Fig. \ref{fig25}), the latter originally claimed to mimic dangling bonds on percolation clusters \cite{r29,r28,r27,r15}.
The study of random walks on a comb structure was initiated as a simple geometrical explanation for anomalous sub-diffusion.  For an infinite comb the resulting diffusion is of the continuous time random walk (CTRW) class. 
We consider ramified structures \cite{r19,r3}, where the comb's backbone is attached to various types of teeth, such as fractals \cite{r10,r12} (see Fig. \ref{fig21}).

Diffusion is measured along the backbone, the $x$ coordinate. 
The particle's sticking times in the teeth are power law distributed provided that the teeth are infinite. This is related to a well known feature of first passage times $\tau$ of ordinary one dimensional random walks in half space, which follows power law statistics. More precisely the probability density function (PDF) of times in the teeth follows $\psi(\tau) \sim \tau^{-\frac{3}{2}}$, $\tau \gg 1$ if the teeth are infinite linear chains (see Fig. \ref{fig25}). For simple comb structures the resulting diffusion is slower than normal, $\langle x^2 \rangle \sim t^{\frac{1}{2}}$, where $t$ is the total measurement time.

Here we investigate distributions of occupation times, i.e. the total time a particle occupies a domain of interest. The occupation times, also called residence times, are related to phenomena such as the total time of phase persistence \cite{r26}, the total number of photons a molecule or a blinking quantum dot emits \cite{r17,r21,r442} and the average total time a particle occupies a region \cite{r1216}. 
Occupation time analysis on combs is relevant to the analysis of the reaction time or rate in a subspace of a comb-like system \cite{r1214}. Single particle tracking allows to measure the statistics with high precision \cite{r1215,r1219}. 
For a short time the dynamics of a finite comb should resemble that of an infinite comb (see details below). We expect to find an occupation time statistics similar to that found in the theory of Weak Ergodicity Breaking (WEB) \cite{r4,r4442,r441,r6}, e.g. the bimodal Lamperti distribution that characterizes CTRW. For a large enough comb this statistics should persist for several time scales until the particle reaches the finite boundaries of the comb. Obviously, a cutoff will follow as the comb is fully explored and we shall obtain the ergodic phase. Such cutoffs have been observed in experiments of fluorescence statistics of nanoparticles \cite{r1218}, surface diffusion on an Ag(100) substrate \cite{r1219} and analytical models \cite{r1217}.  Our work shows the relation between the statistics of the first passage time (FPT) on a single tooth, a cluster of teeth and the occupation time. Thus, we develop a method to compute first passage time statistics on a comb.
The CTRW systems that were examined to this day are finite and have a unique equilibrium, i.e. the occupation time distribution reaches equilibrium. 
One new element  in this research is to check the convergence to asymptotic results and to find the transition times between different behaviors of the system. We find quasi-stationary states, which depend on the finite size of teeth and backbone, and give estimates on transitions between the different states. We also treat the biased comb, which exhibits behaviors different than the unbiased case.

We consider two classes of problems, which we call simply and non-simply connected problems. For the first class we divide the graph into two domains with a single transition path from one part of space to the other through a single point. We ask what is the total time a particle occupies the domain of interest. Here we can use renewal theory to compute the distribution of occupation times. This is made possible after we compute the first passage time statistics on the comb structure. On simply connected domains we find Lamperti statistics, similar to occupation time statistics found for CTRW dynamics \cite{r4}. A more challenging case, are occupation times on non-simply connected domains with multiple connections (entries and exits) to the domain under investigation \cite{r444}. Unfortunately here we do not have exact solutions. Thus, simulations and mean-field theory are used to analyze these problems.
We find non-trivial occupation time statistics for different types of non-simply connected domains. This class of non-simply connected models exhibits statistics different from those obtained with a simpler renewal approach. 

\section{The comb model\label{general_comb_model}}
\begin{figure}
\centerline{\psfig{figure=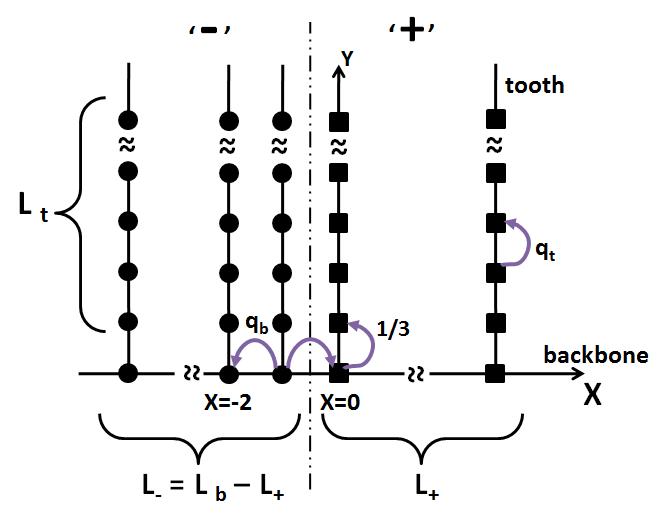,width=90mm,height=65mm}}
\caption{ 
The comb is made of a backbone (the $x$ axis) and teeth. 
The comb is divided into two regions, $'+'$ and $'-'$ with a single boundary between them (dashed dotted line). Occupation times in state $'+'$ are an example for a simply connected problem.}
\label{fig25}
\end{figure}

\begin{figure}
\centering{
\subfigure[]{\centerline{\psfig{figure=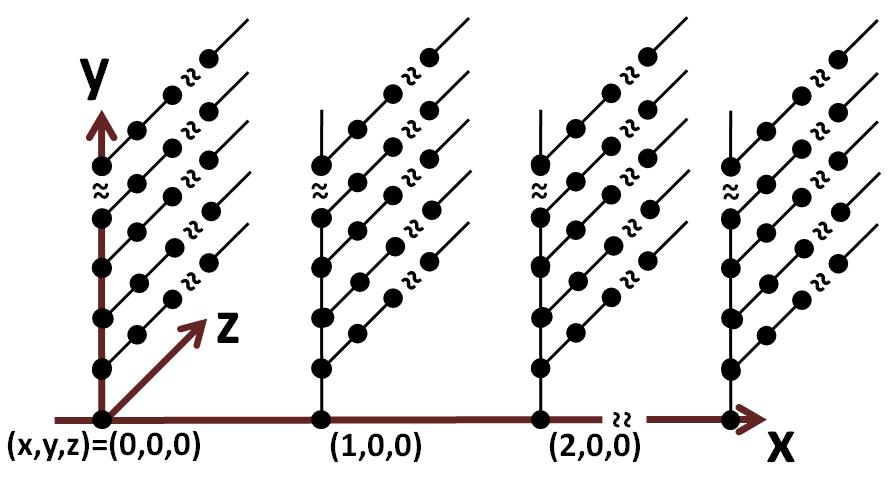,width=90mm,height=45mm}}}

\subfigure[]
{\centerline{\psfig{figure=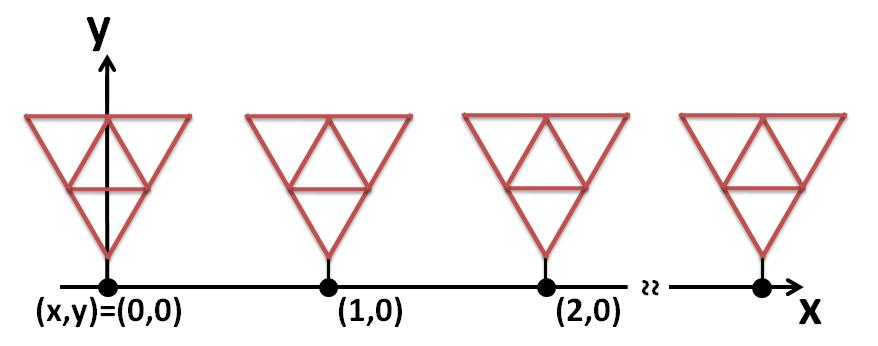,width=90mm,height=45mm}}}

\caption{ Two examples of combs with ramified teeth. (a) An example of a three dimensional comb section with ramified  teeth, is plotted in $(x,y,z)$ space. The backbone is embedded on the $x$ axis. The definition of the ramified structure is given in Sec. \ref{ram_teeth}. (b) An example of a comb with ramified teeth. The teeth are Sierpinski gaskets connected by a linear chain, the backbone. The particle performs a random walk with delays in the teeth. The finiteness of the teeth corresponds to a power law distributed first passage time PDF which is followed by an exponential cutoff at large FPT \cite{r2}.\label{fig21}}}

\end{figure}

Diffusion on infinite long comb structures has been widely studied \cite{r10,r2} since the resulting diffusion is anomalous. The comb contains a backbone and teeth on which ordinary random walk is performed. The backbone and teeth are composed of sites. The teeth stem from the backbone sites. The particle occupies each site for a constant period $\tau_0=1$ before jumping to nearest neighbors. For now we assume that each tooth is a one dimensional linear chain. The backbone and teeth lengths are $L_b$ and $L_t$ respectively (see Fig. \ref{fig25}). A particle performs a discrete time random walk on the comb hopping to one of its nearest neighbors with transition probabilities soon to be specified.

The hopping probability on the backbone sites to the right neighbor is $q_b^+$ and $q_b^-$ to the left neighbor. These are given by
\begin{equation}
q_b^\pm = \frac{1  \mp \epsilon}{3},
\label{eq65}
\end{equation}
where $|\epsilon| < 1$ and the drift $\epsilon>0$ directs the movement towards the $'-'$ region. The probability to hop up to the tooth is $1/3$. 
The random walk performed on the teeth is unbiased with a transition probability up or down equal to $q_t = 1/2$. The tooth's extreme site is a reflecting boundary. 
The case of $\epsilon=0$ is the unbiased random walk on the comb.

In what follows,  we consider the occupation time of the random walk in a given region. That is the total time a particle spends in some domain for a random walk taking time $t$. We consider different domains in which the occupation time is calculated. We will start with a simply connected domain. 

\subsection{Occupation time on a simply connected domain}

We divide the comb's backbone into two regions $'+', x\geq0$, and $'-',x<0$. Each has a backbone length of $L_+$ and $L_- = L_b- L_+$ respectively. 
The boundary is perpendicular to the $x$ axis, located between points $(x,y)=(0,0)$ and $(-1,0)$ (see Fig. \ref{fig25}).
Our goal is to find the occupation time fraction in space $'+'$, i.e., the time the particle occupies the region $(x\geq0)$ over the total measurement time. At time $n=0$ the particle starts the random walk on a site near the boundary.

\begin{figure}

\centerline{\psfig{figure=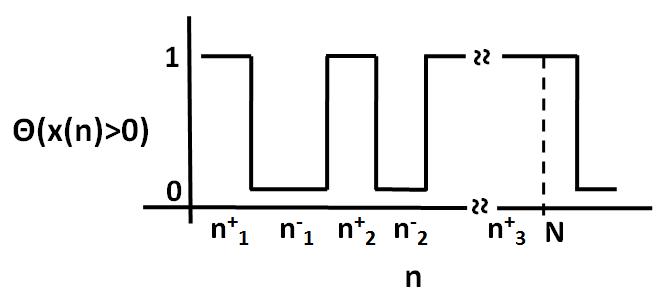,width=90mm,height=65mm}}

\caption{ 
A two state renewal process. The particle jumps between the $'+'$ and $'-'$ regions of the comb. The time intervals $n_i^+$ and $n_i^-$ are the waiting times in states $'+'$ and $'-'$ respectively. We assign values of $1 (0)$ according to the occupation of the $'+' ('-')$ region respectively. The total measurement time is $N$.}
\label{fig26}
\end{figure}

For a particle starting on $(0,0)$ let $n^+$ be the number of steps until the particle reaches $(-1,0)$ for the first time ($n^+$ is a discrete first passage time in state $'+'$). $n^-$ has a similar meaning. The renewal sequence $(n_1^+,n_1^-,n_2^+,n_2^-,\ldots)$ describes the dynamics in states $'+'$ and $'-'$ (see Fig. \ref{fig26}). We define a two state renewal process on the comb, where the particle jumps between states $'+'$ and $'-'$. This process will be useful later.
The observable
\begin{equation}
T_+ (N) =\sum_{n=1}^N {\Theta(x(n)>0)}
\label{eq59}
\end{equation} 
is the total occupation time of the $'+'$ region, where $\Theta(\cdot)$ is the Heaviside step function, $N$ the total measurement time and $x(n)$ is the x-axis coordinate at time $n$.
The occupation time fraction of the $'+'$ region is
\begin{equation}
\bar{p}_+ = \frac{T_+ (N)}{N} . 
\label{eq25}
\end{equation} 

The first passage time on a segment is the first time a particle exits it \cite{r443,r445,r446}.
We now relate between statistics of the first passage time on one tooth with the PDF of the FPT on a region $'+'$, using a method developed in \cite{r2}. The FPT on a tooth is the time a particle reaches $(x,0)$ starting from $(x,1)$. The discrete FPT probability function (PF) of a tooth for $n \ll L_t^2$ is given by
\begin{equation}
F_{L_t \to \infty} (n)= \left\{
\begin{array}{l l}
\frac{ \Gamma(n/2)}{2\Gamma(\frac{1}{2})  \Gamma (\frac{n+3}{2})}\Bigg| _{n \gg 1} \sim {\sqrt{\frac{2}{\pi}} \text{ }  n^{-\frac{3}{2}}}  & \text{ if n is odd}, \\
 0  & \text{ if n is even}.
\end{array}
\right.
\label{eq15}
\end{equation}
We notice the famous $n^{-\frac{3}{2}}$ dependence in the long time limit \cite{r2} (see some details in appendix \ref{Comb_derv}).
After a long enough time, $n \gg L_t^2$, the particle reaches the reflecting boundary which causes an exponential cutoff,
\begin{widetext}
\begin{equation}
F_{L_t} (n)\Big| _{n \gg L_t^2} \sim \left\{
\begin{array}{l l}
\frac{ 2}{{L_t}}  \sin^2 (\frac{\pi}{2{L_t}})  \cos^{n-1} (\frac{\pi}{2L_t})\Big| _{L_t \gg 1} \sim \frac{\pi^2}{2  {L_t}^3} \text{ }  e^{-\frac{\pi^2}{8  {L_t}^2 } (n-1)} & \text {   if n is odd}, \\
 0 & \text {  if n is even},  
\end{array}
\right.
\label{eq16}
\end{equation}
\end{widetext}
which is derived in appendix \ref{Comb_derv}. 
An exact expression for $F_{L_t} (n)$ can be found in principle from the inversion of $\hat{F}_{L_t} (z)$. The Z-transform $\hat{F}_{L_t} (z)= \sum_{n = 0}^{\infty} {F_{L_t}(n) \text{ } z^n}$,  is a useful tool and is given by Eq. (\ref{eq32}). Let $\hat{F}_{L_\pm,L_t}^\pm (z)$ be the Z transform of the discrete FPT PF $F_{L_\pm,L_t}^\pm (n^\pm)$ for the time in the $'\pm'$ region which depends of course on $L_\pm$ and $L_t$.  In appendix \ref{comb_FPT_PDF} we find
\begin{widetext}
\begin{equation}
\hat{F}_{L_\pm,L_t}^\pm (z)=\sqrt{\frac{q_b^\mp}{q_b^\pm}} \frac{ q_b^\mp \sinh[L_\pm \hat{\phi}^c (z)] - q_b^\pm  \sinh[(L_{\pm} -2) \hat{\phi}^c (z)]}
{ q_b^\mp \sinh[(L_{\pm} +1)  \hat{\phi}^c (z)] - q_b^\pm \sinh[(L_{\pm}-1) \hat{\phi}^c (z)]},
\label{eq29}
\end{equation}
\end{widetext}
where
\begin{equation}
\cosh(\hat{\phi}^c (z)) = \frac{\hat{w}_{L_t} (z)}{2 z \sqrt{q_b^+ q_b^-}}
\label{eq73}
\end{equation}
and
\begin{equation}
\hat{w}_{L_t} (z)=1-\frac{z}{3} \hat{F}_{L_t} (z).
\label{eq731}
\end{equation}
$\hat{w}_{L_t} (z)$ is called the weighted delay time polynomial of the particle inside each tooth \cite{r2}. For further details see appendix \ref{comb_FPT_PDF}. 
%
Eq. (\ref{eq29}) relates between the first passage time on a single tooth Eq. (\ref{eq32}) and the first passage time on a domain with $L_+$ or $L_-$ teeth.
We notice two special cases of Eq. (\ref{eq29}). The first is the form of $\hat{F}_{L_+,L_t} (z)$ in the unbiased case, $\epsilon=0$,
\begin{equation}
\hat{F}_{L_+,L_t}(z)=\frac{\cosh{[({L_+}-1) \hat{\phi}^c (z)]}}{\cosh{[{L_+} \hat{\phi}^c (z)]}}.
\label{eq79}
\end{equation}
The second case is that of the infinite long section $L_+ \to \infty$,
\begin{equation}
\hat{F}_{L_+ \to \infty,L_t}(z) = e^{-\hat{\phi}^c (z) }=\frac{\hat{w}_{L_t} (z)}{2 z \sqrt{q_b^+ q_b^-}}-\sqrt{\frac{\hat{w}_{L_t}^2 (z)}{4 z^2 q_b^+ q_b^- }-1} .
\label{eq803}
\end{equation}

We switch to the continuum limit, where time is a continuous variable (instead of $n$). Namely, we consider the occupation fraction
\begin{equation}
\bar{p}_+ = T_+ /t,
\label{eq566}
\end{equation} 
where $t$ is the total measurement time.
To do so, we switch from the Z transform to the Laplace transform applying a technique from \cite{r4}, taking $z= e ^{-u  \tau_0}$, where $u$ is the standard Laplace variable and $\tau_0=1$ is the constant time step between consecutive jumps. We now briefly review the theory of occupation time for two state renewal processes \cite{r26,r21,r4}.

\subsection{Renewal process statistics\label{renewal}}
 
The system jumps back and forth between two states $'+'$ and $'-'$. The $i^{\text{th}}$ sojourn time spent on each region is described by the PDFs
\begin{equation}
F^\pm (\tau) \sim \frac{A^\pm}{|\Gamma(-\alpha^\pm)|}   \tau^ {-(1+\alpha^\pm)} 
\label{eq39}
\end{equation} 
where $\tau$ is large and $0< \alpha^\pm<1$.  In this case the averaged times in states $'+'$ and $'-'$ diverge. 

The power law behavior of the corresponding Laplace transforms is
%
\begin{equation}
\hat{F}^\pm (u) = \int^{\infty}_{0} {F^\pm (\tau) e^{-u \tau} d \tau} \sim 1- A^\pm  u^{\alpha^{\pm}}
\label{eq07}
\end{equation} 
in the small $u$ limit. The total occupation time is the sum of the sojourn times spent in the $'+'$ state.
We are only interested in the long time limit result.
If $\alpha^+$ and $\alpha^-$ differ, the particle is stuck in the state with smaller $\alpha$ in the long time limit.
The occupation time PDF turns into a Dirac delta function.
If the two $\alpha^\pm$ are equal, $\alpha^\pm=\alpha$, then the coefficients $A^\pm$ become relevant.

The Lamperti PDF, here denoted as $\delta_{\alpha,Q} (\bar{p}_+)$, describes the PDF of the occupation time fraction, Eq. (\ref{eq566}), for such a renewal process \cite{r26,r44}.  It is a natural generalization of 
the arcsine distribution. It has appeared in several applications \cite{r4,r17,r18,r441,r181,r182,r183}
\begin{widetext}
\begin{equation}
\delta_{\alpha,Q} (\bar{p}_+) = \frac{\sin{(\pi \alpha)}}{\pi}  \frac{Q (1-\bar{p}_+)^{\alpha-1}  \bar{p}_+^{\text{ }\alpha-1}} {\bar{p}_+^{2 \alpha} + Q^2  (1-\bar{p}_+)^{2 \alpha} + 2  Q (1-\bar{p}_+)^{\alpha}  \bar{p}_+^{\text{ }\alpha} \cos{(\pi  \alpha)}}.
\label{eq08}
\end{equation}
\end{widetext}
$\alpha=\alpha^\pm$  is the power law exponent given in Eq.  (\ref{eq07}). We define $Q$, the asymmetry parameter
\begin{equation}
Q =  \frac{A^+}{A^-}.
\label{eq09}
\end{equation}
The special case of $\alpha=1/2$ and $Q=1$  is  the well known arcsine law.
Notice the two divergences of Eq. (\ref{eq08}) on $\bar{p}_+ = 0, 1$. This happens since the particle gets stuck at one of the states for a duration which is comparable to the total
 measurement time. The divergences disappear for $\alpha=1$ as the Lamperti PDF
becomes a Dirac delta function.
It's first and second cumulants are:
\begin{equation}
\langle \bar{p}_+ \rangle = \frac{Q}{1+Q} = \frac{A^+}{A^- + A^+}
\label{eq11}
\end{equation}
and
\begin{equation}
\begin{array}{l l}
\sigma^2_{p_+}=\langle \bar{p}_+^2 \rangle - \langle \bar{p}_+ \rangle^2 =  \\
\text{ }\text{ }\text{ }\text{ }\text{ }\text{ }\text{ }\text{ }\text{ } (1-\alpha) \langle \bar{p}_+ \rangle   (1- \langle \bar{p}_+ \rangle)= (1-\alpha) \frac {Q}{(1+Q)^2}. 
\end{array}
\label{eq12}
\end{equation}
%
%
%

\subsection{Unbiased comb\label{unb_comb}}
\subsubsection{The infinite long backbone and teeth\label{unb_inf_comb}}

We now consider the infinite comb with infinite long teeth and backbone \cite{r10,r2,r16} $L_t,L_\pm \to \infty$. We first consider the unbiased case $\epsilon = 0$. 
Taking the long time limit or $u \to 0$, we find the tooth's
asymptotic FPT PDF in Laplace space using  Eq. (\ref{eq13})
\begin{equation}
\hat{F}_{L_t \to \infty} (u) \sim  1 - \sqrt{2u}.  
\label{eq35}
\end{equation}
The entire region's FPT generating function, the continuous analogue of Eq. (\ref{eq29}), is found using Eqs. (\ref{eq731},\ref{eq803},\ref{eq35}). The generating function is
\begin{equation}
\hat{F}_{L_\pm \to \infty,L_t \to \infty}^{\pm} (u) \sim  1 - ({2  u})^\frac{1}{4}, 
\label{eq33}
\end{equation}
hence in this case 
\begin{equation}
F_{L_\pm \to \infty,L_t \to \infty}^{\pm} (\tau) \sim  2^\frac{1}{4} \Big|\Gamma \Big(-\frac{1}{4} \Big)\Big|^{-1 } \tau^{-\frac{5}{4}}. 
\label{eq74}
\end{equation}
Although the domains $'+'$ and $'-'$ are infinite, the particle always returns with probability $1$ to the boundary site. In particular, from Eqs. (\ref{eq07},\ref{eq08},\ref{eq33}) we have $\alpha^\pm=1/4$ and $Q=1$. Hence using Eq. (\ref{eq08})
the occupation time statistics has the form of a Lamperti function
\begin{equation}
\delta_{\frac{1}{4},1} (\bar{p}_+) = \frac{1}{\sqrt{2} \pi}\frac{[(1-\bar{p}_+)  \bar{p}_+]^{-\frac{3}{4}}}
{\sqrt{\bar{p}} + \sqrt{1-\bar{p}_+} + \sqrt{2} \text{ } [(1-\bar{p}_+) 
 \bar{p}_+]^{\frac{1}{4}} }.
\label{eq30}
\end{equation}
%

\subsubsection{Infinite backbone and finite teeth\label{3D3}}

We consider an infinite comb with an infinite backbone and finite teeth.
At short times, the particle doesn't ``feel'' the finiteness of the teeth. Thus, according to Eq. (\ref{eq55}), for short times $t \ll L_t^2$,
The system behaves as an infinite comb with Eqs. (\ref{eq35}-\ref{eq30}) being valid. The occupation fraction's statistics is given by Eq. (\ref{eq30}).
Once $t \gg L_t^2$, the generating function of the tooth's FPT PDF turns to
\begin{equation}
\hat{F}_{L_t} (u) \sim  1 -2 \Big( L_t-\frac{1}{2} \Big) \text{ } u,
\label{eq36}
\end{equation}
as the finiteness of the teeth takes effect. In other words the small $u$ expansion is now analytical unlike Eq. (\ref{eq35}) and the average FPT is finite. The region's FPT PDF is derived from the generating function using  Eqs. (\ref{eq731},\ref{eq803},\ref{eq36})
\begin{equation}
\hat	{F}_{L_\pm \to \infty,L_t}^{\pm} (u)\sim 1 - \sqrt{2 (1+L_t) u}.
\label{eq37}
\end{equation}
It exhibits a non-analytical behavior since $L_\pm \to \infty$. See subsection \ref{ramified_tooth} for further discussion. Thus now 
\begin{equation}
F_{L_\pm \to \infty,L_t}^{\pm} (\tau) \sim  \sqrt{\frac{1+L_t}{2\pi}}  \tau^{-\frac{3}{2}}. 
\label{eq75}
\end{equation}
Comparing with Eq. (\ref{eq07}) we see that $A_+=A_-$ and $\alpha_+=\alpha_-=1/2$.
The occupation fraction PDF is found from Eq. (\ref{eq08}),
\begin{equation}
\delta_{\frac{1}{2},1} (\bar{p}_+) =  \frac{1}{\pi}  \frac{ 1}{ \sqrt {(1-\bar{p}_+)  \bar{p}_+} } 
\label{eq56}
\end{equation}
which is the famous arcsine law. Since the teeth are finite, the diffusion is effectively one dimensional. Hence we get L\'evy's well known result for statistics of occupation times in half plane for one dimensional Brownian motion.

\subsubsection{finite backbone and infinite teeth\label{3D4}}

We now consider a comb model with infinite long teeth and a finite backbone of length $L_b$. 
We divide it into two regions of backbone lengths $L_+$ and $L_-$ 
(see Fig. \ref{fig25}).
The sojourn waiting time PDF of each tooth is determined using Eq. (\ref{eq35}). 
As the particle begins its movement near the boundary site (0,0) it doesn't sense the finiteness of the backbone, provided that $L_{+}$ and $L_{-}$ are both much larger than unity. The emerging statistics is that of the infinite backbone and teeth model and the occupation time fraction PDF is given by Eq. (\ref{eq30}), $\delta_{\frac{1}{4},1} (\bar{p}_+)$, for short (but not very short) times.

Figures \ref{fig13} and \ref{fig15} depict simulations. Regarding Fig. \ref{fig13} with $L_{+}=L_{-}$, for not too long times the system attains a quasi-stationary state. This state is simply the case of the comb of infinite backbone and teeth, found for time scales when the particle did not have time to reach the reflecting boundaries. To calculate the short time variance of $\bar{p}_+$ we use Eq. (\ref{eq12}), where $\alpha=1/4$ and $Q=1$, to find $\sigma^2_{p_+}=3/16=0.1875$ accordingly.  The first quasi-stationary state exists for several time scales. The particle reaches one of the reflecting boundary sites of the backbone at time $t_{f} \sim \frac{\pi}{2}  \text{ } \min{(L_+^4,L_-^4)}$ (see further details in subsection \ref{trans_scales}). 
After this transition time the backbone is no longer effectively infinite. The long time equilibrium (the final plateau in Fig. \ref{fig13}) is found using $Q=1$ due to symmetry and $\alpha=1/2$ due to the excursions into the infinite long teeth.  Eq. (\ref{eq12}) gives $\sigma^2_{p_+}=1/8=0.125$ and the occupation time statistics is  $\delta_{\frac{1}{2},1} (\bar{p}_+)$. In Fig. \ref{fig13} the transition from  $\delta_{\frac{1}{4},1} (\bar{p}_+)$, in short times, to {$\delta_{\frac{1}{2},1}(\bar{p}_+)$, in the long times, behaves in qualitative agreement with  $t_{f}$ as the backbone length $L_b$ is varied.
\begin{figure}

\centerline{\psfig{figure=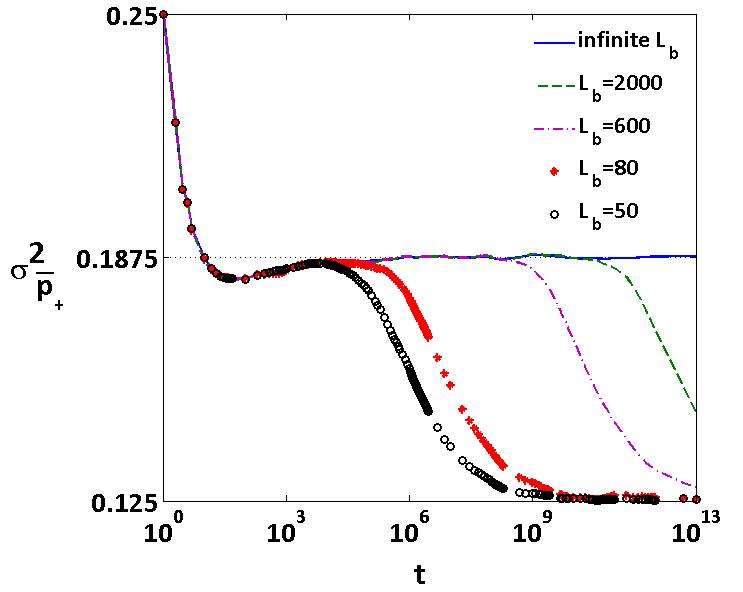,width=90mm,height=65mm}}
\caption{ 
The variance of $\bar{p}_+$ versus time $t$ where $\bar{p}_+$ is the occupation time fraction in half space. The comb has infinite long teeth and a backbone with varying size illustrated in the figure. After a transition time which depends on $L_b$, the number of sites on the backbone, the particle ``feels'' the finiteness of the backbone and transits from a quasi-equilibrium state to the final equilibrium. The variance for short times is that of an infinite backbone and teeth model, $\sigma^2_{\bar{p}_+}= 0.1875$. At long times, the equilibrium for the finite backbone and infinite teeth model has a variance $\sigma^2_{\bar{p}_+}= 0.125$. Averages over 20,000 trials were used in the simulation.}
\label{fig13}
\end{figure}
\begin{figure}
\centerline{\psfig{figure=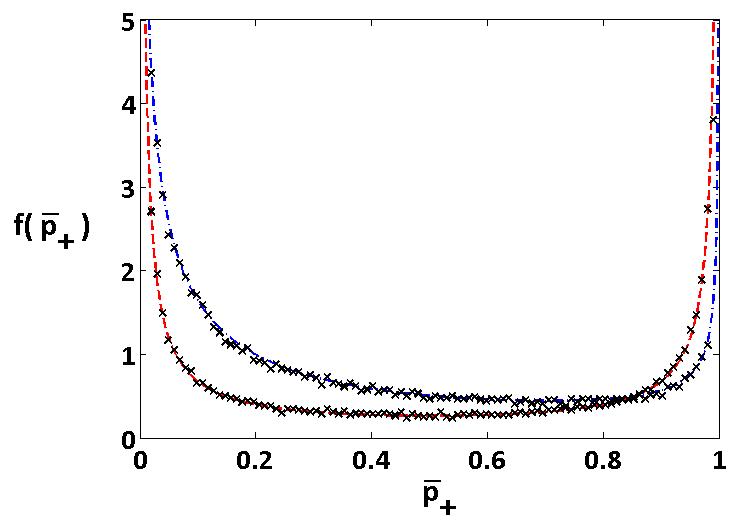,width=90mm,height=65mm}}

\caption{ 
The PDF of occupation fraction for a comb system with finite backbone and infinite teeth, simulations (crosses) versus theory (curves). The region of interest is $L_+=25$ for a backbone of length $L_b=75$.  For short times ($t=4 \times 10^{3}$),  the occupation time statistics behaves as if the system has infinite teeth and backbone. It's PDF is given by $\delta_{\frac{1}{4},1} (\bar{p}_+)$, Eq. (\ref{eq30}) (the dashed curve). Afterwards the occupation fraction PDF converges to it's final equilibrium $\delta_{\frac{1}{2}, 0.495} (\bar{p}_+)$, Eq. (\ref{eq57}) (the dotted dashed curve). This PDF is depicted in the figure at $t=10^{12}$. Notice the asymmetry in the final state due to the fact that $L_+ \neq L_-$. Averages over 110,000 trials were used in the simulation.}
\label{fig15}
\end{figure}

Let us explain the numerical results presented in Fig. \ref{fig15}. In the figure we see that the PDF of $\bar{p}_+$ is non-symmetric for long measurement times. In the general case, after $t_{f}$, the Laplace transform of the FPT PDF in regions $'\pm'$ is found using Eq. (\ref{eq29})
\begin{equation}
\hat{F}_{L_\pm,L_t \to \infty}^{\pm} (u) \sim  1 - 2\Big(L_\pm -\frac{1}{2}\Big) \sqrt{ \frac{u}{2}},
\label{eq38}
\end{equation}
where the subtraction of the half stems from the reflecting boundary condition (derivation of this result is similar to the derivation of Eq. (\ref{eq36})). The multiplication factor by two is explained by the back and forth movement of the particle on the backbone. Here in this case $\alpha=1/2$ and if $L_+ \neq L_-$ we have asymmetry. The statistics of the occupation fraction is now given using Eqs. (\ref{eq08}, \ref{eq38}), 
\begin{equation}
\lim_{t \to \infty} \delta_{\frac{1}{2},\frac{ L_+ - \frac{1}{2}}{ L_- -\frac{1}{2}}} (\bar{p}_+) = \frac{1}{\pi} \frac{\frac{ L_+ - \frac{1}{2}}{ L_- -\frac{1}{2}}  [(1-\bar{p}_+)  \bar{p}_+]^{-\frac{1}{2}}}
{ \bar{p}_+ + \Big(\frac{ L_+ - \frac{1}{2}}{ L_- -\frac{1}{2}}\Big)^2  (1-\bar{p}_+) }.
\label{eq57}
\end{equation}
A similar result is also found in \cite{r4} with a one dimensional CTRW model. The ensemble average of the occupation fraction is
\begin{equation}
\langle \bar{p}_+ \rangle = \frac{L_+ - \frac{1}{2}}{L_+ + L_- - 1} .
\label{eq62}
\end{equation}
In Fig. \ref{fig15} we portray the occupation fraction PDF of a system with $L_b=75, L_+=25$ and $L_t \to \infty$ for short and long times. For short times, the occupation time statistics behaves as if the system has infinite teeth and backbone, as expected. At long times it sets in a stationary state of finite backbone and infinite teeth with $\delta_{\frac{1}{2}, 0.495} (\bar{p}_+)$ according to Eq. (\ref{eq57}).

\subsection{Not simply connected spaces}
\subsubsection{Periodic partitioning\label{per_div}}

Up until now we divided the system into two regions, $'+'$ and $'-'$. Importantly, these regions are connected by a single link between sites. In these cases, applying renewal theory leads to exact expressions for large times. For occupation time statistics one can of course divide the system in many other ways. One example is shown in Fig. \ref{fig9}, where a comb with infinite backbone and teeth is considered. We define two types of teeth, those with squares and those with circles. In each unit cell (circumvented by a rectangle) we have $L_{p+}$ teeth with squares and $L_{p-}$ teeth with circles, where $L_\pm \to \infty$. We consider the occupation fraction on the teeth with squares, the $'+'$ region (see Fig. \ref{fig9}). In this example $L_{p+}=2$ and $L_{p-}=3$. Here there is no single boundary site separating the $'+'$ and $'-'$, as in previous cases. Still, we can use the symmetry and map the problem onto a finite system, which backbone length is $L_{p+} + L_{p-}$ teeth, with periodic boundary conditions at the exterior backbone sites (see Fig. \ref{fig22}). The system is divided into two regions of $L_{p+}$ and $L_{p-}$ teeth. The FPT PDF of region $'+'$ is defined as the earliest time for a particle on the exterior teeth of the backbone domain $'+'$ to exit to the $'-'$ domain. In the small $u$ limit Eq. (\ref{eq29}) gives
\begin{figure}
\centerline{\psfig{figure=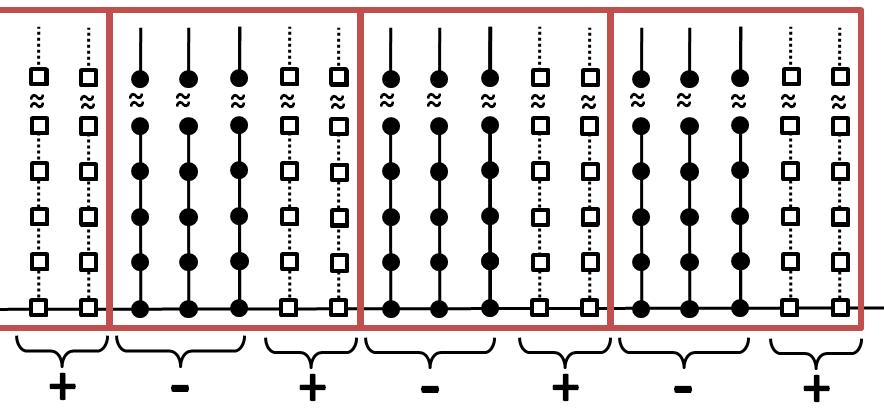,width=90mm,height=45mm}}
\caption{ 
An infinite comb structure with periodic partitioning. The teeth are divided periodically into unit cells (circumvented by rectangles), containing $L_{p+} + L_{p-} =5$ teeth. The region of interest is made up of all the teeth with squares, $L_{p+} =2$ teeth in each unit cell.}
\label{fig9}
\end{figure}
\begin{equation}
\hat{F}_{L_{p\pm},L_t \to \infty}^{\pm} (u) \sim  1 -  L_{p\pm} \sqrt{ \frac{u}{2}}.
\label{eq64}
\end{equation}
Hence $\alpha=1/2$ as expected. More interestingly, Eq. (\ref{eq64}) differs from Eq. (\ref{eq38}) due to the periodic boundary conditions at the two exterior backbone sites. The PDF of the occupation time fraction is found using Eq. (\ref{eq08})
\begin{equation}
\delta_{\frac{1}{2},\frac{L_{p+}}{L_{p-}}} (\bar{p}_+) = \frac{1}{\pi}  \frac{\frac{L_{p+}}{L_{p-}}  [(1-\bar{p}_+)  \bar{p}_+]^{-\frac{1}{2}}}
{ \bar{p}_+ + \Big(\frac{L_{p+}}{L_{p-}}\Big)^2  (1-\bar{p}_+) }.
\label{eq10}
\end{equation}
The FPT on the teeth are responsible for the exponent $\alpha=1/2$ in the Lamperti PDF Eq. (\ref{eq10}). The average occupation probability is found using Eq. (\ref{eq11})
\begin{equation}
\langle \bar{p}_+ \rangle = \frac{L_{p+}}{L_{p+} + L_{p-}} .
\label{eq63}
\end{equation}
In Sec. \ref{3D4}, as $L_\pm \gg 1$ Eqs. (\ref{eq57},\ref{eq62}) converge to Eqs. (\ref{eq10},\ref{eq63}) respectively since the boundary conditions turn out uninstrumental for our observable.
\begin{figure}
\centerline{\psfig{figure=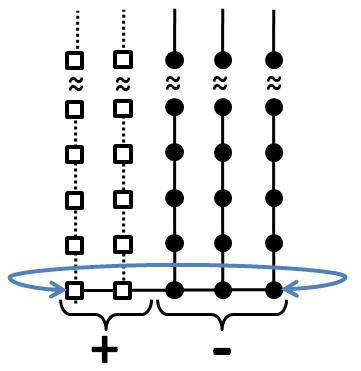,width=40mm,height=40mm}}
\caption{ 
A comb structure with periodic boundary conditions. The comb has a finite long backbone and infinite long teeth. The $'+'$ region of interest are the two teeth with squares. $L_{p+}=2, L_{p-}=3$ in this example.}
\label{fig22}
\end{figure}

\subsubsection{Unbalanced periodic partitioning}

We now analyze a more challenging case. Consider the occupation time on odd positive $x$ (see Fig. \ref{fig10}),
\begin{equation}
t_{\text{odd } x>0} = \sum_{k=0}^{\infty} {\int_0^t {d\tau \text{ } \Theta(x(\tau) =2k+1)}}.
\label{eq636}
\end{equation}
Our goal is to find the occupation time fraction statistics on this domain. 

We define the total occupation time on the $x>0$ region as $t_{x>0}$ and $t$ as the total measurement time. The PDF of $\bar{p}_{x>0}=t_{x>0} / t$ is given by Eq. (\ref{eq30}), 
\begin{equation}
F(\bar{p}_{x>0}) =\delta_{\frac{1}{4},1} (\bar{p}_{x>0}). 
\label{eq76}
\end{equation}
The total time in the region of interest is $t_{\text{odd }x>0}$. 
Now we return to the occupation fraction. We use the equation
\begin{equation}
\frac{t_{\text{odd }x>0}}{t} = \frac{t_{x>0}}{t} \text{  } \frac{t_{\text{odd }x>0}}{t_{x>0}}
\label{eq40}
\end{equation}
as a starting point for the solution. Notice that ${t_{x>0}}/{t}$ and ${t_{\text{odd }x>0}}/{t_{x>0}}$ have a known Lamperti distribution. This equation holds in its probabilistic counterpart 
\begin{equation}
P[{\text{odd }x>0}]) = P[{x>0}] \text{  } P[{\text{odd}|x>0}].
\label{eq41}
\end{equation}
$P[{x>0}]$ is the probability of occupying the positive x-axis and $P[{\text{odd}|x>0}]$ is the conditional probability of occupying the odd sites of x-axis while staying at $x>0$. Our goal of interest is to find the PDF of $\overline{P[{\text{odd }x>0}]}={t_{\text{odd }x>0}}/{t} $.

The statistics of $\overline{P[{\text{odd}|x>0}]}$ are given by the result in Sec. \ref{per_div} with $m=n=1$, i.e., its PDF is in the long time limit
\begin{equation}
F(\text{  } \overline{P[{\text{odd}|x>0}]} \text{  }) = \delta_{\frac{1}{2},1} (\text{  } \overline{P[{\text{odd}|x>0}]}\text{  }).
\label{eq42}
\end{equation}
The random variables $\overline{P[{\text{odd}|x>0}]}$ and $\overline{P[x>0]}$ are correlated. To treat the problem approximately we use a mean-field (MF) approach. Namely, we neglect the correlations between the  $\overline{P[{\text{odd}|x>0}]}$ and $\overline{P[x>0]}$ random variables. Since the PDFs of $\overline{P[x>0]}$ and $\overline{P[{\text{odd}|x>0}]}$ are known, Eq. (\ref{eq76}) and Eq. (\ref{eq42}), respectively, we find the PDF of our observable within MF approximation.
\begin{figure}
\centerline{\psfig{figure=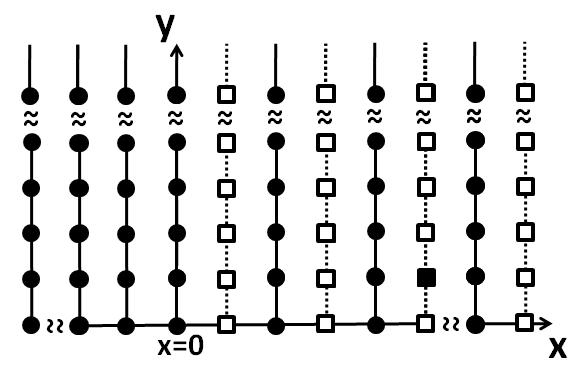,width=65mm,height=45mm}}
\caption{ 
The occupation time of the non-simply connected infinite comb. We consider occupation times in the region $x>0$ and $x$ being odd using a mean-field approach. The region of 
interest is marked by squares.}
\label{fig10}
\end{figure}
\begin{widetext}
$$\delta_{\text{odd }x>0}^{\text{MF}} (\overline{P[{\text{odd }x>0}]}) = $$
$$\int_0^1  {d \overline{P[{x>0}]}   \int_0^1 
 {d \overline{P[{\text{odd}|x>0}]} \text{ }  \delta_{\frac{1}{4},1} (\overline{P[{x>0}]}) \text{ } \delta_{\frac{1}{2},1} 
(\overline{P[{\text{odd}|x>0}]}) \text{ } \delta ( \overline{P[{\text{odd }x>0}]} - \overline{P[{x>0}]} \text{  } \overline{P[{\text{odd}|x>0}]})
}}=$$
\begin{equation}
\int_0^1 \frac{d \overline{P[{x>0}]}}{\overline{P[{x>0}]}} \text{ }  \delta_{\frac{1}{4},1} (\overline{P[{x>0}]}) \text{ }
\delta_{\frac{1}{2},1} \Bigg( \frac{\overline{P[{\text{odd }x>0}]}}{\overline{P[{x>0}]}}\Bigg).
\label{eq21}
\end{equation}
\end{widetext}
The cumulants of the MF approximation can be calculated analytically, 
$\langle \overline{P[{\text{odd }x>0}]}^{\text{MF}} \rangle = 1/4$ (which is exact) and $\langle (\overline{P[{\text{odd }x>0}]}^{\text{MF}})^2 \rangle - 
\langle \overline{P[{\text{odd }x>0}]}^{\text{MF}} \rangle^2=\frac{13}{128} = 0.102..$. The above was compared to a numerical simulation for an infinite unbiased comb with a particle initiated next to the boundary between the regions. Our numerics show $\langle \overline{P[{\text{odd }x>0}]} \rangle=1/4$ as expected and 
\begin{equation}
\begin{array}{l l}
\sigma^2_{\text{odd }x>0}= \\
\\
\text{ }\text{ }\text{ }\text{ }\text{ } \langle \overline{P[{\text{odd }x>0}]}^2\rangle-\langle\overline{P[{\text{odd }x>0}]}\rangle^2 =0.109..
\end{array}
\label{eq210}
\end{equation} 
The following figures depict the statistics (see Fig. \ref{fig11}) and the time evolution of the variance (see Fig. \ref{fig12}).  
\begin{figure}
\centerline{\psfig{figure=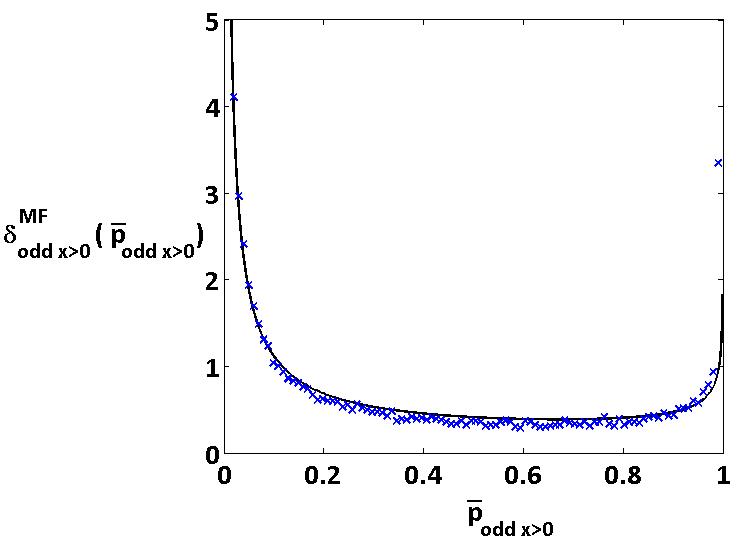,width=90mm,height=65mm}}
\caption{ 
PDF of occupation time fraction for the non-simply connected domain in Fig. \ref{fig10}, simulations (crosses) versus theory (curve). The occupation times are calculated on the odd numbered sites $x>0$. The occupation time fraction PDF is approximated by Eq. (\ref{eq21}) (the solid curve). The simulation graph was calculated at a long measurement time $t=10^{13}$. Over 44,000 systems were used in the simulation. MF theory seems to work well.}
\label{fig11}
\end{figure}
\begin{figure}
\centerline{\psfig{figure=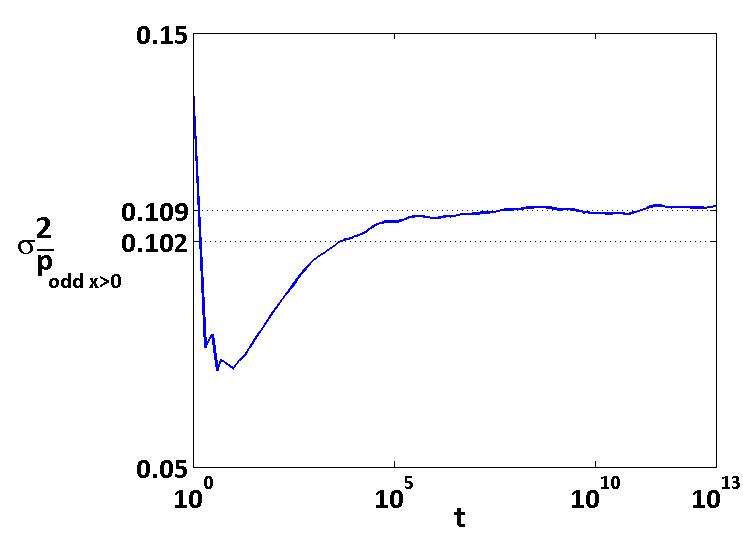,width=90mm,height=65mm}}
\caption{ 
The variance of the occupation time fraction versus measurement time for a non-simply connected system on the infinite comb. The region of interest are the odd numbered sites $x>0$. The $\sigma^2_{\text{odd }x>0}$ as a function of the measurement time converges to $0.109..$. Also shown is the MF prediction of the variance $0.102...$ . 44,000 random walks on a comb were used in the simulation.}
\label{fig12}
\end{figure}

\subsection{biased comb}

We now turn on a bias in the backbone, a constant drift to the left $\epsilon>0$. The random walk on the teeth remains unbiased. This is analogous to a flow through a structure with a main channel (the backbone) \cite{r27} or a charged particle diffusing in a field aligned with the backbone.

\subsubsection{Biased infinite backbone and infinite teeth \label{bias_inf}}
 
We consider the occupation fraction in $x>0$, namely a simply connected system. For a biased process with $\epsilon>0$, the particle will eventually end up in region $'-'$. Hence the occupation fraction in state $'+'$ is trivial in the long time limit.  Namely, the PDF of $\bar{p}_+$ is a delta function $\delta(\bar{p}_+)$, so  $\bar{p}_+=0$. For short times and a small bias the particle isn't affected by the bias (when it starts close to the boundary between domains). Then  $\bar{p}_+$ has a PDF like the unbiased case of $\delta_{\frac{1}{4},1 (\bar{p}_+)}$, Eq. (\ref{eq30}).  
This transition is shown in Fig. \ref{fig18} where we plot $\sigma^2_{\bar{p}_+}$ versus time. For short times, Eq. (\ref{eq12}) gives $\sigma^2_{\bar{p}_+} = 0.1875$ since $\alpha={1}/{4}$ and $Q=1$.
\begin{figure}
\centerline{\psfig{figure=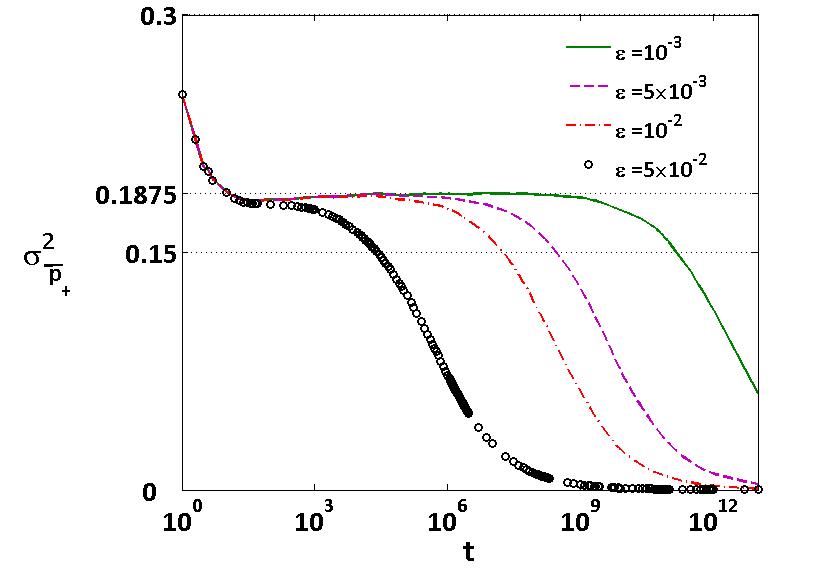,width=90mm,height=65mm}}

\caption{ 
The infinite comb system $L_\pm, L_t \to \infty$, with a biased backbone $ \epsilon >0 $. The occupation time fraction exhibits the behavior of an infinite comb for time 
$t \ll (2\pi  \epsilon^4)^{-1}$ and $ |\epsilon| \ll 1$.
The occupation time fraction PDF is $\delta_{\frac{1}{4},1} (\bar{p}_+)$ for short times. At $t \gg ({2\pi  \epsilon^4})^{-1}$, $\sigma^2_{\bar{p}_+} \to 0$. Averages over 50,000 trials.}
\label{fig18}
\end{figure}
For long times we have $\sigma^2_{\bar{p}_+} = 0$, since as mentioned, the bias drives the particle to region $'-'$ (see Fig. \ref{fig18}). As expected, the transition time, between these two behaviors, increases as $|\epsilon|$ decreases. Let us estimate this transition time. In subsection \ref{trans_scales} we show that the mean square displacement is (see Eq. (\ref{eq53})), 
\begin{equation}
\langle x^2 (t) \rangle \Big| _{t \to \infty}  \approx 4 \epsilon^2 \text{ } t +2\sqrt{\frac{2t}{\pi}}.
\label{eq60}
\end{equation}
When $\epsilon=0$, $\langle x^2 (t) \rangle \Big| _{t \to \infty}  \sim \sqrt{t} $ is well known. The first term $O(t)$ in Eq. (\ref{eq60}) is larger then the second $O(\sqrt{t}
)$ when $t>t_{d}$,
\begin{equation}
t_{d} \sim \frac{1}{2\pi \text{ }\epsilon^4} .
\label{eq61}
\end{equation}
We see a very long transition time that decays like $\epsilon^{-4}$. Eq. (\ref{eq61}) is in qualitative agreement with our numerical simulation (see Fig. \ref{fig18}). Here $t_{d1}=2.5 \times 10^4$, $t_{d2}=1.6 \times 10^{7}$, $t_{d3}=2.5 \times 10^8$ and $t_{d4}=1.6 \times 10^{11}$ for $\epsilon_1=5 \times 10^{-2}$, $\epsilon_2=10^{-2}$, $\epsilon_3=5 \times 10^{-3}$ and $\epsilon_4=10^{-3}$ respectively. 

\subsubsection{Biased finite backbone and infinite teeth\label{FBIT}}

We consider a system containing a finite backbone and infinite teeth. The FPT PDF of the teeth is determined by the generating function Eq. (\ref{eq35}). The FPT PDF of the $'+'$ and $'-'$  regions is given in Laplace space using Eq. (\ref{eq29}). In the small $u$ limit
\begin{equation}
\hat{F}_{L_\pm,L_t \to \infty} (u) \sim 1-\frac{1}{|\epsilon|} \Bigg|1- (1 \mp \epsilon) \Bigg( \frac{1 \mp \epsilon}{1 \pm \epsilon} \Bigg)^{L_\pm-1} \Bigg| \Big( \frac{u}{2} \Big)^{\frac{1}{2}}.
\label{eq43}
\end{equation}
Using renewal theory, Sec. \ref{renewal}, the occupation time PDF is found to be the Lamperti function $\delta_{\frac{1}{2},Q} (\bar{p}_+)$ with
\begin{equation}
Q=\Bigg| \frac{ 1- (1 - \epsilon) \Big( \frac{1 - \epsilon}{1 + \epsilon} \Big)^{L_+-1}}{1- (1 + \epsilon) \Big( \frac{1 + \epsilon}{1 - \epsilon} \Big)^{L_- -1}}\Bigg|
\label{eq44}
\end{equation}
and
\begin{widetext}
\begin{equation}
\sigma^2_{\bar{p}_+} =
\frac{ |((1 + \epsilon) ^{L_+ -1} - (1 - \epsilon)^{L_+ }) 
((1 + \epsilon)^{L_- } - (1 - \epsilon) ^{L_- -1})| }{2 [ (1+\epsilon)^{L_+ + L_- -1} - (1-\epsilon)^{L_+ + L_- -1}]^2} {(1-\epsilon)^{L_- -1} (1+\epsilon)^{L_+ -1}}.
\label{eq77}
\end{equation}
\end{widetext}
In Fig. \ref{fig19} we show $\sigma^2_{\bar{p}_+}$ versus time with one transition between two quasi-stationary states. The first step is to hop to the $'+'$ domain with probability $\frac{1}{2}$. Thus at $t \to 0$, $\langle \bar{p}_+ \rangle = {1}/{2}$ and $\sigma^2_{\bar{p}_+} = {1}/{4}$. The system then settles in a quasi-stationary diffusion state which corresponds to an unbiased infinite backbone and teeth system with $\sigma^2_{\bar{p}_+} = 0.1875$. Here the bias is unimportant as explained already. At a longer time we see a transition to the behavior described by Eq. (\ref{eq77}), i.e. $\sigma^2_{\bar{p}_+} = 0.0802$ or $\sigma^2_{\bar{p}_+} = 0.0514$ for $\epsilon=0.01$ or $\epsilon=0.05$ respectively. In principle we may have three quasi-equilibrium states, namely, two transitions. This may happen if $|\epsilon|$ is very small such that $|\epsilon| \ll \sqrt{{2}\text{ }{\pi}^{-1}} \text{ } {\max({L_+,L_-})}^{-1}$. The quasi-equilibrium states are: (a) A diffusive phase, where $\epsilon,L_b$ are non-important and the PDF of occupation is $\delta_{\frac{1}{4},1} (\bar{p}_+)$. (b) Since $\epsilon$ is very small, particles ``feel'' the finiteness of the backbone $L_b$ (but not the bias) and Eq. (\ref{eq57}) is valid. (c) The bias kicks in and $\sigma^2_{\bar{p}_+}$ converges to Eq. (\ref{eq77}).

\begin{figure}
\centerline{\psfig{figure=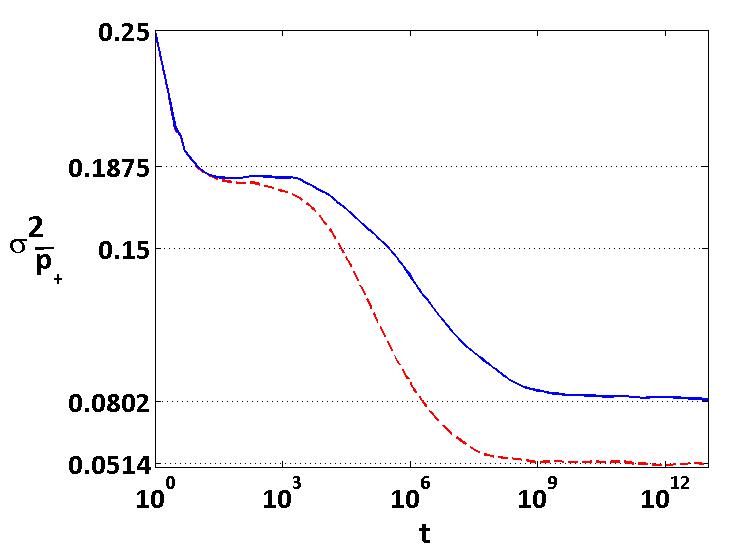,width=90mm,height=65mm}}


\caption{ 
The comb system with a finite biased backbone and infinite long teeth. The cases of interest are $L_+ = 202$ sites out of $L_b=224$ with $\epsilon=0.05$ (the solid curve) and $L_+=90$ sites out of $L_b=100$ with $\epsilon=0.01$ (the dashed curve).  $\sigma^2_{\bar{p}_+}$ exhibits a shoulder close to $\sigma^2_{\bar{p}_+} = 0.1875$  which is the value for the unbiased infinite comb. The occupation time fraction PDF converges to the Lamperti PDF with  $\alpha=\frac{1}{2}$ and $Q$ according to Eq. (\ref{eq44}). Averaging over 64,000 comb random walks are made.}
\label{fig19}
\end{figure}

\subsubsection{Finite backbone and finite teeth with bias\label{FBFTB}}

Finally we consider a finite system. The FPT PDF of the teeth is given in Laplace space by Eq. (\ref{eq36}). We take the small $u$ limit of Eq. (\ref{eq29}) to find the FPT PDF generating functions of the $'+'$ and $'-'$ regions   
\begin{equation}
\hat{F}_{L_\pm,L_t} (u) \sim 1-\frac{L_t+1}{{|\epsilon|}} \Bigg|1- (1 \mp \epsilon) \Bigg( \frac{1 \mp \epsilon}{1 \pm \epsilon} \Bigg)^{L_\pm-1} \Bigg| u.
\label{eq49}
\end{equation}
So the average times in $'+'$ and $'-'$ are finite.
Its occupation fraction PDF is a Lamperti $\delta_{1,Q} (\bar{p}_+) $ with $Q$ given in Eq. (\ref{eq44}). This is a delta function centered on
\begin{equation}
\langle \bar{p}_+ \rangle =
 (1-\epsilon)^{L_- -1} \Bigg|\frac{(1 + \epsilon) ^{L_+ -1} - (1 - \epsilon)^{L_+  } }{ (1+\epsilon)^{L_+ + L_- -1} - (1-\epsilon)^{L_+ + L_- -1}}\Bigg|.
\label{eq78}
\end{equation}
This result is valid in the long time limit after the particle has explored the finiteness of both backbone and teeth. It is expected from the the ergodic hypothesis.

\section{Ramified teeth}

\subsection{A definition of ramified teeth \label{ram_teeth}}

Up until now we examined a tooth comprised of a one dimensional linear chain of sites (see Fig. \ref{fig25}), which we call a regular tooth. 
We now generalize the analysis for a simply connected problem on a comb with ramified teeth. That means we replace the teeth with fractal objects. See Fig. \ref{fig21} and details below.
We are interested in structures that have a power law FPT PDF \cite{r443}. Some examples are
fractals such as the Sierpinski gasket \cite{r3,r10,r12,r2} and T fractal \cite{r3,r2}.
Other ramified structures are given in \cite{r24}.

\subsection{The FPT exponent of an exactly decimable fractal}

Diverging from our main scope we remind the reader of a family of fractals, which FPT exponent is directly related to the local spectral dimension $\tilde{d}_s$ of the fractal \cite{r19,r3,r2}. Exactly decimable fractals \cite{r3} are defined as decimation invariant if it is possible to eliminate a subset of points (and all the bonds connecting these points) obtaining a network with the same geometry of the starting one. 
The spectral dimension is defined by
\begin{equation}
\tilde{d}_s=2 \frac{\tilde{d}_f}{\tilde{d}_w},
\label{eq98}
\end{equation} 
where $\tilde{d}_f$ is the fractal dimension and $\tilde{d}_w$ is the scaling exponent of the random walk. As a reminder, a mass $m$ encircled by a hypersphere of radius $r$ scales as $m \propto r^{\tilde{d}_f}$. The time $t$ it takes to transverse a distance $r$ scales as $t \propto r^{\tilde{d}_w}$. 
Notice that for the euclidean lattice $\tilde{d}_f=d$ is the euclidean dimension and $\tilde{d}_w=2$ for a simple Gaussian random walk. Thus the local spectral dimension coincides with the euclidean dimension $\tilde{d}_s=d$. 

The first passage PDF behaves like $F(t) \sim t^{-(1+\alpha)}$ in the long time limit.
According to \cite{r10,r2}, for $\tilde{d}_s<2$ the FPT exponent $\alpha$ is given by
\begin{equation}
\alpha = 1-\frac{\tilde{d}_s}{2}.
\label{eq100}
\end{equation} 
The last equation is the main connection between the geometry and the occupation time statistics of the system. This holds for renewal processes, namely for simply connected domains. The next section links between the exponent $\alpha$ and the expression for the occupation time statistics.

\subsection{Occupation time statistics on a comb\label{ramified_tooth}}

We present a simple method for calculating the asymptotic behavior
of the first passage time PDF of the unbiased ramified comb section.
We consider two separate cases: infinite and finite long backbone of combs with ramified teeth. We use the discrete Eq. (\ref{eq29}) to obtain an expression for the 
FPT PDF of the ramified comb of infinite backbone segment length $L_+ \to \infty$. First, we remind the reader that the sojourn time in a tooth has the same asymptotic dependence on time as the FPT out of the tooth. Thus the generating functions of the sojourn time, $\hat{\Psi}_{tooth} (u)$, and the FPT, $\hat{F}_{tooth} (u)$, are related by
\begin{equation}
1- \hat{\Psi}_{tooth} (u) \propto 1- \hat{F}_{tooth} (u) \propto u^\alpha,
\label{eq50}
\end{equation}
where $0 < \alpha \leq 1$.
Using the same method of \cite{r4} we transform the problem from the discrete time formalism to continuous time $t$. We do so in Laplace space by inserting $z=\hat{\Psi}_{tooth} (u)$ in Eq. (\ref{eq29}). Instead of using the first passage of the tooth we replace it with an ``effective'' single site by setting $\hat{w}_{L_t} (z)= 2/3$ in Eq. (\ref{eq731}). For a ramified comb with infinite backbone $L_\pm \to \infty$ , Eq. (\ref{eq803}) yields in the small $u$ limit
\begin{equation}
1 - \hat{F}^{\text{ ram comb}}_{L_+ \to \infty} (u) \propto u^\frac{\alpha}{2}.
\label{eq19}
\end{equation}
Now we have the tools to predict the occupation time statistics of the ramified comb using Eq. (\ref{eq08})
\begin{equation}
F( \bar{p}_+) =  \delta_{\frac{\alpha}{2},1} (\bar{p}_+).
\label{eq99}
\end{equation}
For an unbiased dynamics of the ramified comb with $L_\pm \to \infty$ we find a new exponent half the size of the exponent of the ramified tooth's FPT PDF. For regular teeth of infinite length $\alpha=1/2$ and we get Eq. (\ref{eq30}).  

Consider a specific type of ramified tooth with $d$ generations, which is constructed by the following rules. For $d=1$ the tooth is simply an infinite linear chain (like in the original comb model). For $d=2$, we use the linear chain ($d=1$ object) and from each site ``grow" a linear chain (see Fig. \ref{fig21}a). Such objects can be extended to generation $d$. The ramified comb has a FPT generating function $\hat{F}^{\text{ ram comb}}_{L_b \to \infty} (u) \sim 1- B_d  \sqrt[2^{d+1}]{u}$ in the small $u$ limit according to Eq. (\ref{eq19}). So in this example $\alpha=2^{-d}$. If the tooth is finite, then this lasts for several time scales. Following the effect of the particle reaching a boundary at a certain generation, the FPT PDF Laplace transform evolves to $1-\hat{F}^{\text{ ram comb}}_{L_b \to \infty} (u) \propto \sqrt[2^{d}]{u}$. This effect occurs if the particle reaches a boundary of either the extent of the comb's ramified teeth or the finite backbone. The transition of the FPT PDF corresponds to the transition of the occupation fraction statistics Eq. (\ref{eq08}), where $\alpha$ coincides with the exponent of the FPT PDF. Other related analyses are given in \cite{r10,r16}.

For a finite backbone segment we use the tools from \cite{r4} and Sec. \ref{3D4},\ref{per_div} and \ref{FBIT}. For an unbiased ramified comb domain with a finite backbone segment (and a reflecting boundary condition at the end) we get
\begin{equation}
1 - \hat{F}^{\text{ ram comb}}_{L_+} (u) \propto 
2  \Big(L_+ -\frac{1}{2}\Big) u^{\alpha} .
\label{eq97}
\end{equation}
The occupation fraction PDF is the Lamperti function Eq. (\ref{eq08}), $\delta_{\alpha, Q} (\bar{p}_+)$ with
\begin{equation}
Q = \frac{2 L_+ -1}{2 L_- -1}. 
\label{eq101}
\end{equation}
We now turn to explore the transition times between quasi-stationary statistics which prevail for several time scales.
 
\section{Transition times\label{trans_scales}}

Here we estimate transition times of the dynamics, namely, the time scales when the particles start ``feeling'' the boundaries. These transitions depend on the dimensions of the system and the value of the bias $\epsilon$. We use a one dimensional CTRW on a lattice \cite{r10,r25} as a tool to predict these transitions. In this model jumps are made to nearest neighbors with a transition probability to the right $(1-\epsilon)/2$ and $|\epsilon| \leq 1$. The motion is unbiased for $\epsilon=0$. The sojourn time power law PDF is given by
\begin{equation}
\psi (\tau) \Big| _{\tau \to \infty} \sim \frac{A}{|\Gamma( -\alpha)|} \text{ } \tau ^ {-(1+\alpha)} 
\label{eq01}
\end{equation} 
where $\tau > 0$ is the sojourn time, $A>0$ and $0< \alpha < 1$.  In this case $\langle\tau\rangle$, the statistical average, diverges.  
The Laplace transform of the sojourn time PDF is given in the small $u$ limit	
\begin{equation}
\hat{\psi} (u)\sim 1 - A u^\alpha.
\label{eq52}
\end{equation}
The mean square displacement is used to estimate transition times. For CTRW it is given in Laplace space by \cite{r10},
\begin{equation}
\langle \hat{x}^2 (s) \rangle = \frac{2 \epsilon^2}{s} \left[ \frac{\hat{\psi} (s)}{ 1-\hat{\psi} (s)} \right] ^2 +\frac{\hat{\psi} (s)}{s  [1-\hat{\psi} (s)]},
\label{eq04}
\end{equation} 
where $s$ is the Laplace conjugate of $t$. For the infinite comb mentioned in Sec. \ref{unb_inf_comb},\ref{bias_inf} $\alpha=1/2$ and Eqs. (\ref{eq35},\ref{eq60}) are valid. 
In the time domain, for an open system 
\begin{equation} 
\langle x^2 (t) \rangle \Big| _{t \to \infty}  \approx \frac{2 \epsilon^2  t^{2\alpha}}{ A^2  \Gamma(1+2\alpha)}+\frac{(1-4 \epsilon^2)t^\alpha}{ A  \Gamma(1+\alpha)}.
\label{eq05}
\end{equation} 
Notice that in $\langle x^2 (t) \rangle$ we have two terms which are $\epsilon$ dependent and $\epsilon$ independent. The later describing the fluctuations in the absence of a bias. We consider only the long time limit and $|\epsilon| \ll 1$, thus
\begin{equation} 
\langle x^2 (t) \rangle \Big| _{t \to \infty,|\epsilon| \ll 1} \approx \frac{2 \epsilon^2  t^{2\alpha}}{ A^2  \Gamma(1+2\alpha)}+\frac{t^\alpha}{ A  \Gamma(1+\alpha)}.
\label{eq53}
\end{equation} 
For shorter times the diffusion term (i.e. the second term) is dominant over the drift term. The time scale for this transition occurs when these two terms coincide
\begin{equation}
t_{d} \sim \left[ \frac{  A  \Gamma(1+2\alpha)}
{2  \epsilon^2  \Gamma (1+\alpha)} \right] ^ {\frac{1}{\alpha}}.
\label{eq06}
\end{equation}
Notice that the smaller $\epsilon$ the larger is $t_d$, and for $\alpha=0.5$ we have $t_d \propto \epsilon^{-4}$. This transition behavior is also found for occupation times (numerically), see Sec. \ref{bias_inf}.

We now study systems with finite backbone of length $L_+ + L_-$ (see Fig. \ref{fig25}). The teeth are infinite. When the system reaches the boundary $L_{\text{min}}=\min{(L_+,L_-)}$, we expect a transition in the qualitative behavior of the system. Using Eq. (\ref{eq53}), $\langle x^2 (t_f) \rangle \sim L^2_{\text{min}}$, we find
\begin{equation}
\frac{t_{f}^\alpha}{A}  \sim \min{\Big( \Gamma(1+\alpha) L_{\text{min}}^2,\sqrt{\frac{\Gamma (1+2\alpha)}{2}}  \frac{L_{\text{min}}}{|\epsilon|}\Big)} .
\label{eq54}
\end{equation}
Another transition is found for the FPT of the teeth of length $L_t$ using Eq. (\ref{eq54}) with $\alpha=1, \epsilon=0$ (normal diffusion)
\begin{equation}
t_{f}|_{\alpha=1,\epsilon=0} \sim L_t^2
\label{eq55}
\end{equation}
which is well known.

\section{Summary}

Random walks on partitioned infinite comb give non-trivial occupation time statistics. We found that the method of partitioning dramatically affects the statistics. We examined two main cases: simply and non-simply connected problems. The first were treated analytically as renewal processes. Non-simply connected problems are especially interesting since renewal theory framework does not apply, namely, we have intricate correlations between sojourn times in the domain. The problem was analyzed heuristically with a mean-field approach. The same system partitioned in various ways can fall into either cases of connectedness.  For a non-simply connected problem we suggest extending the mean-field approach to solve various systems, as will be shown in \cite{r262}.

For simply connected problems the particle in our analysis and simulations always starts near the boundary between the two regions. We now assume the contrary, that particle starts far enough from the boundary. The time it takes to reach the boundary may be of the order of time that the particle "feels" the finiteness of the comb or the bias activated on it. The resulting occupation time statistics will differ on account of the initial condition. Another example of the influence of  initial conditions is a comb with finite backbone and long enough teeth, where a particle starts at the edge of the backbone far away from the boundary. Trivially the immediate occupation time statistics will not be $\delta_{\frac{1}{4},1} (\bar{p}_+)$ but a delta function centered on 1 or 0 for $x(t=0) = L_+ -1 $ or $x(t=0) = - L_-$ respectively. Thus our results depend on the initial conditions. Further work is needed to investigate the role of initial conditions.

The random walk on the comb is an approximation for various physical phenomena \cite{r10}. Here we find the occupation time statistics of a particle for various subspaces on a comb system. The occupation time exhibits successive statistics stretching for several time scales. In a real world experiment this understanding is crucial in correctly interpreting the system's composition and evolution.

We find the transition times between quasi-stationary states with different residence times statistics. The transitions are controlled by the measurement time, the value of the bias and the finite dimension of the system (see Eqs. (\ref{eq61},\ref{eq06},\ref{eq54})). For finite CTRW systems the occupation time statistics converges to a Lamperti PDF. Statistics of that kind is also observed in the finite comb model for several decades of time. Finally the dynamics experiences a cutoff and the distribution of the occupation time converges to a delta function. This is of course ergodic behavior. Another example of the transitions of occupation time statistics is for the biased comb with infinite long backbone and teeth, where the duration of the initial Lamperti statistics $\delta_{\frac{1}{4},1} (\bar{p}_+)$ is $t_d \propto \epsilon^{-4}$. 
As mentioned in the introduction occupation times are related to diffusion-influenced reaction \cite{r1214}, hence the dynamical evolution of the system can be viewed through fluorescence quenching in a system, where the traps are concentrated in one of the comb's regions.


\textbf{Acknowledgments:  } This work was supported by the Israeli Science Foundation.

\appendix

\section{FPT of the Tooth\label{Comb_derv}}

The first passage time on a tooth of the comb is the first time that a particle exits the tooth. It can be analyzed using techniques found in \cite{r15,r4,r2}. A single tooth, on $x=1$ for example, is depicted in Fig. \ref{fig25}. We assume the tooth is of length $L_t$. For the calculation an absorbing boundary condition is placed at $(1,0)$ at the edge of the tooth. A particle starts at site $(1,1)$ and travels along the tooth. The other edge of the tooth is a reflecting boundary condition in site $(1,L_t)$. For the discrete time analysis $F_{L_t} (n)$, the discrete FPT probability function (PF), is the probability of exiting the tooth at $n$ for the first time.  The generating function \cite{r2} is the Z-transform
\begin{equation}
\hat{F}_{L_t} (z)= \sum_{n = 0}^{\infty} {F_{L_t}(n) \text{ } z^n}=\frac{\cosh{[({L_t}-1) \hat{\phi} (z)]}}{\cosh{[{L_t} \hat{\phi} (z)]}},
\label{eq32}
\end{equation}
where $\cosh{(\hat{\phi} (z))}= z^{-1},$\text{ }$  |z| \leq 1$ \cite{r2}. A similar derivation will be given in the appendix \ref{comb_FPT_PDF}. For an infinite tooth, $L_t \to \infty$, the generating function is
\begin{equation}
\hat{F}_{L_t \to \infty} (z) = e^{-\hat{\phi} (z)}=\frac{1}{z}-\sqrt{\frac{1}{z^{2}}-1}.
\label{eq13}
\end{equation}
$F_{L_t \to \infty} (n)$ is given by the the coefficients of $z^n$ in the expansion of Eq. (\ref{eq13}), which yields Eq. (\ref{eq15}). 

The solution of the long time limit discrete PDF PF  for the finite tooth involves an Abellian theorem \cite{r15}
\begin{equation}
\lim_{n \to \infty} \frac{F_{L_t} (n)}{G (n)} = \lim_{z \to z_0} \frac{\hat{F}_{L_t} (z)}{\hat{G} (z)}.
\label{eq66}
\end{equation}
$z_0$ is the radius of convergence and $G(n)$ is an auxiliary function which behaves the same as $F_{L_t} (n)$ in the long time limit.  We notice that the poles of Eq. (\ref{eq32}) are $z^p_{j}=\sec{\Big(\frac{\pi}{L_t} \Big( j + \frac{1}{2} \Big)\Big)}$, where $j=1,2,\ldots {L_t}$ and $|z^p_j|>1$. $F_{L_t} (n)$ takes a nonzero value only in odd times, thus only the absolute values of the poles are relevant. The smallest most dominant pole in the long time limit is $z^p_{L_t}= \sec{\Big(\frac{\pi}{2 L_t}\Big)}$ which is also the radius of convergence $z_0$. The  discrete FPT PF takes the asymptotic form of $F_{L_t} (n) \Big|_{\text{odd } n \to \infty} \approx D  \big(z^p_{L_t} \big)^{-n}$. This is the form of the auxiliary function we will use,
\begin{equation}
G (n)= \left\{
\begin{array}{l l}
 \big(z^p_{L_t} \big)^{-n}  & \text{ if } n > 0 \text{ is odd}, \\
 0  & \text{ if n is even}.
\end{array}
\right.
\label{eq67}
\end{equation}
This yields Eq. (\ref{eq16}), which is valid at $n >> 2 L_t^2$. The finite tooth discrete FPT PF can be approximated at short times $n << 2 L_t^2$ by Eq. (\ref{eq15}), a power law. 

\section{A comb section's discrete FPT PF\label{comb_FPT_PDF}}

We examine a comb section with a backbone $L_+$ teeth of length $L_t$. The comb section's FPT $F^+_{L_+,L_t} (n)$ is given by the inverse Z transform of $\hat{F}^+_{L_+,L_t} (z)$, Eq. (\ref{eq29}) (see Fig. \ref{fig25}). 

We calculate the discrete FPT PF of the $'+'$ segment using the method in \cite{r2}. ${P}_{x,y} (n)$ is the probability of occupying the $(x,y)$ cell at time $n$. The particle starts at $(1,0)$, thus $P_{1,0} (0)=1$. The boundary condition at $(0,0)$ is absorbing. We write the master equations:
\begin{widetext}
\begin{equation}
\begin{array}{l l}
F^+_{L_+,L_t} (n+1) = \frac{1+\epsilon}{3} P_{(1,0)} (n), \\
P_{(1,0)} (n+1) = \frac{1+\epsilon}{3} P_{(2,0)} (n) + \frac{1}{3} \sum_{i=0}^{\infty} {F_{L_t} (n-i) P_{( 1,0)} (i)}, \\
P_{(2,0)} (n+1) = \frac{1+\epsilon}{3} P_{(3,0)} (n) +\frac{1-\epsilon}{3} P_{(1,0)} (n) + \frac{1}{3} \sum_{i=0}^{\infty} {F_{L_t} (n-i) P_{(2,0)} (i)},  \\
\ldots  \\
\ldots  \\
P_{(L_+ -2,0)} (n+1) = \frac{1+\epsilon}{3} P_{(L_+ - 1 ,0)} (n) +\frac{1-\epsilon}{3} P_{(L_+ - 3,0)} (n) + \frac{1}{3} \sum_{i=0}^{\infty} {F_{L_t} (n-i) P_{( L_+ -2,0)} (i)},  \\
P_{(L_+ -1,0)} (n+1) = \frac{2}{3} P_{(L_+,0)} (n) +\frac{1-\epsilon}{3} P_{(L_+-2 ,0)} (n) + \frac{1}{3} \sum_{i=0}^{\infty} {F_{L_t} (n-i) P_{( L_+ -1,0)} (i)},  \\
P_{(L_+,0)} (n+1) = \frac{1-\epsilon}{3} P_{(L_+ - 1,0)} (n) +\frac{1}{3} \sum_{i=0}^{\infty} {F_{L_t} (n-i) P_{( L_+,0)} (i)}.
\end{array}
\label{eq68}
\end{equation}
The Z transform gives:
\begin{equation}
\begin{array}{l l}
\hat{F}^+_{L_+,L_t} (z) =q_b^- z \hat{P}_{(1,0)} (z), \\
\hat{P}_{(1,0)} (z) = q_b^- z \hat{P}_{(2,0)} (z) + \frac{1}{3} z  \hat{F}_{L_t} (z) \hat{P}_{(1,0)} (z) + 1, \\
\hat{P}_{(2,0)} (z) = q_b^- z \hat{P}_{(3,0)} (z) +q_b^+ z \hat{P}_{(1,0)} (z) + \frac{1}{3} z  \hat{F}_{L_t} (z) \hat{P}_{(2,0)} (z)),  \\
\ldots  \\
\ldots  \\
\hat{P}_{(L_+ -2,0)} (z) = q_b^- z \hat{P}_{(L_+ - 1 ,0)} (z) +q_b^+ z\hat{P}_{(L_+ - 3,0)} (z) + \frac{1}{3} z  \hat{F}_{L_t} (z) \hat{P}_{(L_+ -2,0)} (z),  \\
\hat{P}_{(L_+ -1,0)} (z) = \frac{2}{3} z \hat{P}_{(L_+,0)} (z) +q_b^+  z\hat{P}_{(L_+-2 ,0)} (z) +  \frac{1}{3} z  \hat{F}_{L_t} (z) \hat{P}_{(L_+ -1,0)} (z),  \\
\hat{P}_{(L_+,0)} (z) = q_b^+ z\hat{P}_{(L_+ - 1,0)} (z) +  \frac{1}{3} z  \hat{F}_{L_t} (z) \hat{P}_{(L_+ ,0)} (z).
\end{array}
\label{eq69}
\end{equation}
\end{widetext}
The terms $\hat{P}_{(x ,0)} (z)$ and $\frac{1}{3} z  \hat{F}_{L_t} (z) \hat{P}_{(x,0)} (z)$ describe the occupation of the the tooth's sites. The later is the resulting arrival of the particle from the tooth itself onto the backbone site. The weighting time polynomial $\hat{w}_{L_t} (z)$, Eq. (\ref{eq731}), is derived from both by its definition. The set of equations above, Eq. \ref{eq69}, can be further simplified:
\begin{equation}
\begin{array}{l l}
\hat{F}^+_{L_+,L_t} (z) =q_b^- z \hat{P}_{(1,0)} (z), \\
\hat{w}_{L_t} (z) \hat{P}_{(1,0)} (z) = 1+q_b^- z \hat{P}_{(2,0)} (z), \\
\hat{w}_{L_t} (z) \hat{P}_{(2,0)} (z) = q_b^- z \hat{P}_{(3,0)} (z) +q_b^+ z \hat{P}_{(1,0)} (z),  \\
\ldots  \\
\ldots  \\
\hat{w}_{L_t} (z) \hat{P}_{(L_+ -2,0)} (z) = q_b^- z \hat{P}_{(L_+ - 1 ,0)} (z) +q_b^+ z  \hat{P}_{(L_+ - 3,0)} (z) ,  \\
\hat{w}_{L_t} (z)\hat{P}_{(L_+ -1,0)} (z) = \frac{2}{3} z \hat{P}_{(L_+,0)} (z) +q_b^+ z \hat{P}_{(L_+-2 ,0)} (z) ,  \\
\hat{w}_{L_t} (z) \hat{P}_{(L_+,0)} (z) = q_b^+ z \hat{P}_{(L_+ - 1,0)} (z) .
\end{array}
\label{eq96}
\end{equation}
%
%
Choosing a heuristic solution of a recurrence relation $\hat{P}_{(x,0)} = \hat{f}_{x-1} \hat{P}_{(x-1,0)}  $ and inserting in Eq. (\ref{eq96}), one finds that:
\begin{equation}
\begin{array}{l l}
\hat{f}_{L_+ -1} (z) =\frac{q_b^+ z}{ \hat{w}_{L_t} (z)}, \\
\hat{f}_{L_+ -2} (z) =\frac{\frac{q_b^+ z}{\hat{w}_{L_t} (z) }}{1-\frac{2 z^2 q_b^+}{3 \hat{w}_{L_t}^2 (z)} }, \\
\hat{f}_{x-1} (z) =\frac{\frac{q_b^+ z}{\hat{w}_{L_t} (z)}}{1-\frac{q_b^- z}{\hat{w}_{L_t} (z)} \hat{f}_{x} (z) }     \text{ for     } x=2 \ldots {L_+ -2}.
\end{array}
\label{eq71}
\end{equation}
Eq. (\ref{eq71}) with the two top equations of Eq. (\ref{eq96}) give
\begin{equation}
\begin{array}{l l}
\hat{P}_{1,0} (z) =\frac{\frac{1}{ \hat{w}_{L_t} (z)}}{{1-\frac{ q_b^{-} z}{\hat{w}_{L_t} (z)}  \hat{f}_{1} (z) }}, \\
\\
\hat{f}_0 (z) \equiv \frac{\frac{q_b^+z}{\hat{w}_{L_t} (z)}}{1-\frac{q_b^- z}{\hat{w}_{L_t} (z)} \hat{f}_{1} (z) }.
\end{array}
\label{eq91}
\end{equation}
Manipulating Eq. (\ref{eq91}) and extending it we get
\begin{equation}
\hat{F}^+_{L_+,L_t} (z) = \frac{q_b^-}{q_b^+} \frac{\frac{q_b^+ z}{ \hat{w}_{L_t} (z)}} {1-\frac{ q_b^{-} z}{\hat{w}_{L_t} (z)}  \hat{f}_{1} (z) } =\frac{q_b^-}{q_b^+} \hat{f}_0 (z).
\label{eq92}
\end{equation}
The solution is obtained by assuming
\begin{equation}
\hat{f}_{x} (z)=\frac{\hat{g}_{x} (z)}{\hat{h}_{x} (z)}.
\label{eq72}
\end{equation}
We solve the recurrence relation obtained from Eq. (\ref{eq72}) and the last equation in Eq. (\ref{eq71}). 
\begin{equation}
\begin{array}{l l}
\hat{g}_{x-1} (z) =\frac{q^+_b z}{\hat{w}_{L_t} (z)}  \hat{h}_{x} (z), \\
\hat{h}_{x-1} (z) =\hat{h}_{x} (z) - \frac{q^-_b z}{\hat{w}_{L_t} (z)}  \hat{g}_{x} (z),  \text{ for     } x=2 \ldots {L_+ -2},\\
\hat{h}_{L_+  -1} (z) =1,\\ 
\hat{h}_{L_+  -2} (z) =1-\frac{ 2  q_b^{+} z^2}{3 \hat{w}_{L_t}^2 (z)} . 
\end{array}
\label{eq93}
\end{equation}
The solution is exponential:
\begin{equation}
\hat{h}_{x} (z) = A_+ \hat{\Lambda}_+^{L_+ -1-x} (z) + A_- \hat{\Lambda}_-^{L_+ -1-x} (z).
\label{eq94}
\end{equation}
Applying Eq. (\ref{eq94}) to the second equation in Eq. (\ref{eq93}) we find that
\begin{equation}
\hat{\Lambda}_\pm (z)= \frac{e^{\pm \hat{\phi}^c (z)}}{2 \cosh{\hat{\phi}^c (z)}}
\label{eq95}
\end{equation}
where the definition of $\hat{\phi}^c (z)$ is given in Eq. (\ref{eq73}).
Continuing the development in Eq. (\ref{eq92}) and (\ref{eq93}) we get Eq. (\ref{eq29}). 



\end{document}